# Piezoelectric generator driven by enhanced light pressure

Ha Young Lee[1], Kyung-Won Lim[1,2], Min Sub Kwak[1,2], Geon-Tae Hwang[2] & Sam Nyung Yi[1*]

**Information gathering and analysis by the Internet of Things (IoT) is operating in fields as diverse as environmental management, industry, and healthcare[1]. This work relies on devices such as small actuators and wireless sensors that need a power supply[2]. As the IoT expands, the demand for sensors will increase rapidly, but this expansion might be limited if each device is to have its own external power source, as this would increase the costs and labour involved with installation and maintenance[3]. Much research has therefore sought to develop energy-harvesting technology (e.g., thermoelectric[4,5], triboelectric[6,7], and piezoelectric[8–10]) that can generate electrical energy from irregular energy resources such as light, vibrations, heat, and electric fields. This work reports a light pressure electric generator (LPEG) that converts enhanced radiation pressure into electrical energy. Its underlying principle of relying on light pressure is different from that of conventional piezoelectric power generation. The generator was fabricated by coating a thin Pb(Zr,Ti)O$_3$ film on a wet-chemical-etched microvoid-shaped structure on the surface of a GaAs wafer and then depositing platinum. The surface plasmon phenomenon of its precious metal and microvoid structure enhanced the radiation pressure sufficiently to stimulate the piezoelectric material. The current, voltage, and charge were 2 μA, 0.04 mV, and 25 pC, respectively. This method provides versatile and cost-effective energy harvesting, and represents a potentially useful new way to power the IoT devices.**

A promising unconventional energy source, usually dissipated as waste energy, is piezoelectric generation, which converts vibrations and pressure into electrical energy. Its advantages include high power density and wide applicability compared with other technologies[11]. Suitable vibrations are commonplace: human motion[12], objects' vibrations[13], and sound waves[14] can all provide energy, making piezoelectric generation a potentially efficient, simple, and inexpensive way of powering various IoT devices. However, some environments do not have the necessary movements or vibrations. An alternative energy source that can always be obtained in daily life is light. Solar cells represent the first technology that harvests electrical energy from the environment. However, they face the disadvantage of the manufacturing of solar cells being complicated, and the choice of materials is restricted to those with an energy band gap smaller than that of the light wavelength. An alternative way to harvest light energy uses radiation pressure to stimulate a piezoelectric material whereby the pressure exerted on a surface is caused by the exchange of momentum between it and the electromagnetic field of incident light. However, the radiation pressure of sunlight on Earth, about 4.6 μPa, is too weak to be useful. This greatly limits the potential for light to generate electricity in a piezoelectric material, and amplification is necessary if light is to be used as the only energy source.

The interactions and correlations between metal structures and light wavelength have been studied with various purposes, including the development of optical tunable filters[15], extraordinary transmission[16,17], and optical amplification[18,19]. Research into optical amplification has studied the relationship between metal structures and wavelength in devices such as optical tweezers[20] and optical antennae[21]. These studies have used the principle that the light intensity increases as a result of the surface plasmon phenomenon of light. However, fabrication of a microstructure device for optical amplification has difficulties in post-processing, such as coating a piezoelectric material and electrode processing, and is also expensive.

Therefore, we developed a light-condensing microvoid structure that can collect light in three dimensions and amplify it via surface plasmon phenomena. Fabrication of the structure was achieved by inexpensive wet-chemical etching a GaAs wafer and then building a LPEG with a metal–insulator–metal (Pt–Pb(Zr,Ti)O$_3$–Pt/Ti) structure. We discovered that a thin Pb(Zr,Ti)O$_3$ film can convert radiation pressure from energy irradiated from a laser into electricity, which is a new finding. Finite-difference time-domain (FDTD) simulation (Lumerical Inc.) and the COMSOL Multiphysics program (COMSOL Inc.) analysed the electric field and output voltage in the actual structure. This report of the electricity generation from radiation pressure using a structure of metal and Pb(Zr,Ti)O$_3$ provides a mechanism that is potentially useful to the development of the IoT and furthers our understanding of the natural phenomenon whereby light, metal, and dielectric material work together to make energy.

Fig. 1(a) depicts the sample developed here, showing that power is generated when light passes through the microvoid. Wet-chemical etching the GaAs (100) wafer forms a volcanic-crater-shaped structure with an elliptical frame structure on its top surface. This acts as a convex lens to condense light. The light intensity is further increased by the platinum surface in the microvoid, which forms a surface plasmon. We have experimentally demonstrated this concept for allowing light to stimulate a piezoelectric material. The time for the laser to reach the sample was controlled using a chopper, and measurements were performed at room temperature. Fig. 1(b) and (c) shows the scheme and a photograph of the measurement procedure. The inset of Fig. 1(c) represents a top view of the completed sample comprising a metal–insulator–metal (Pt–Pb(Zr,Ti)O$_3$–Pt/Ti) structure on a wet-chemical-etched GaAs wafer. The upper platinum was patterned to a 1 mm diameter, which was able to sufficiently contain the microvoid.

From the many available methods for selecting or fabricating microvoids, we chose etching to create a volcanic-crater-shaped structure. Fig. 2(a) depicts a microvoid made by wet-chemical etching a single GaAs (100) crystal; it effectively gathers light as shown in Extended Data Videos 1 and 2. The field emission–scanning electron microscopy (FE–SEM) images in Fig. 2(b)–(d) show an etched GaAs microvoid structure coated with electrode and Pb(Zr,Ti)O$_3$. Fig. 2(e) outlines the principles of electricity generation from light pressure. The incident laser photons apply radiation pressure when they hit a surface. The radiation pressure, $P$, is given by $P = 2I/c = 2E^2/c^2\mu_0$, where $I$ is intensity, $c$ is the velocity of light, $E$ is the electric field, and $\mu_0$ is the permeability of free space. For the 10 mW laser used in this experiment, $P = 4.24 \times 10^{-5}$ Pa, and was too weak to induce piezoelectric current. However, amplification of the radiation pressure inside the microvoid then allowed stimulation of the piezoelectric material. First, as the laser hits the metal surface, a surface plasmon is formed and the light intensity increases. The increased intensity of the light then applies a force to the surface metal, and the resulting radiation pressure is transferred to the internal piezoelectric material. This pressure polarizes the piezoelectric material and generates electricity. Fig. 2(f) shows the perovskite structure deformed by radiation pressure. This is fundamentally the piezoelectric effect, but the pressure source is amplified light.

For comparison, the current and voltage were measured for samples otherwise similar to that in Fig. 2 but containing amorphous $SiO_2$ instead of the piezoelectric material or lacking the microvoid. The results of the former comparison case show no jump in current when the laser passed the microvoid (Fig. 3(a)) because the amorphous $SiO_2$ is not a piezoelectric material. The latter case employing a piezoelectric $Pb(Zr,Ti)O_3$ layer in a flat structure (Fig. 3(b)) also showed no jump in output current, as the radiation pressure with the 10 mW laser was insufficient to stimulate the piezoelectric material. However, clear jumps of 2 μA (see Extended Data Video 3) and 0.04 mV were observed when the structure incorporated both a microvoid shape and $Pb(Zr,Ti)O_3$ (Fig. 3(c) and (d)). In addition, the charge amplifier gave 25 pC (Extended Data Fig. 1 and see Extended Data Video 4). These results are consistent with the proposed principle in Fig. 2(e) and clarify the cause of output generation.

Simulations further characterised the trends of the enhanced electric field and the potential distribution in the three-dimensional microvoid structure on the (x–y) and (y–z) planes. The structure and conditions of the FDTD simulation are shown in Extended Data Figs. 2 and 3. The FDTD analysis is fundamentally based on Maxwell's equations, thus allowing the electromagnetic field properties of complex structures to be studied. The FDTD results for the electric-field distribution in Fig. 4 confirm that light resonates and amplifies in the microvoid structure (Fig. 4(b) and (c)), compared with a flat structure (Fig. 4(a)). The 2D COMSOL simulation was based on the calculation of radiation pressure through the behaviour of light inside the microvoid and the electric field from the FDTD data. The direction and range of radiation pressure were applied with reference to the FDTD data (Extended Data Fig. 4), and the magnitude of radiation pressure was applied by converting the value of the electric field acting on the surface. Fig. 4(d)–(f) shows the greater potential of the microvoid structure compared with the flat structure. These data strongly support the results shown in Fig. 3. The potential distribution for the entire structure in the (*x–y*) and (*y–z*) planes is shown in Extended Data Figs. 5 and 6, respectively.

This work proposes a new light-based method for inducing piezoelectric current in $Pb(Zr,Ti)O_3$. Although various generators have used this material, no prior research has demonstrated its stimulation by light (photons). Daylight or common lasers (1 W, ~nN) do not usually have sufficient radiation pressure to stimulate piezoelectric materials, but the resonance and amplification of light demonstrated here using an etched GaAs microvoid allow its conversion to piezocurrent. A single microvoid structure of several hundreds of micrometers obtained a power of 2 μA, 0.04 mV, and 25 pC from a low incident laser power of 10 mW. Optimization of the size, number, and shape of the microvoid structure might increase the output, making this structure potentially very useful. In particular, a LPEG using this principle could power wireless sensors and portable electronics as a new class of self-powering technology that harvests electricity from the environment.

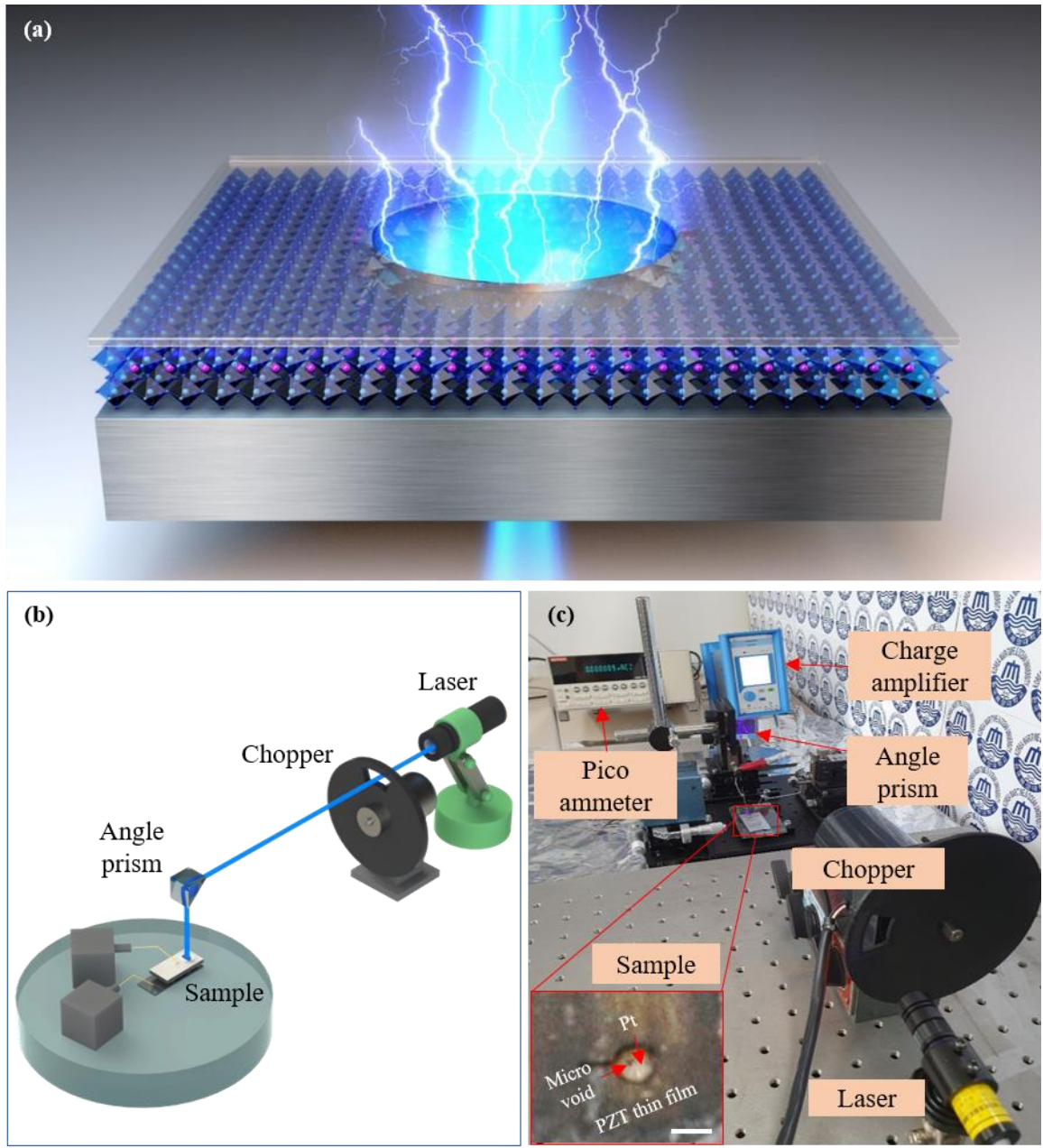

**Fig. 1 | Electricity generation based on enhanced radiation pressure. a**, Concept of energy output from a deformed perovskite structure that increases the intensity of light passing into the microvoid. The GaAs wafer (the metallic block) was wet-chemical etched, and the microvoid had a slit at the bottom to allow the light to pass through. Platinum covering the inside of the microvoid simultaneously acts as the upper electrode and enables surface plasmon formation. **b**, Measurement set-up. The laser passes through a prism to become incident to the microvoid in the sample; current, voltage, and charge are measured using a gold probe. **c**, Photograph of device configuration used during measurement. The laser (401 nm, 10 mW), chopper (~ 54 Hz) and prism are aligned to allow incident light to hit the sample's microvoid. The inset in Fig. 1 (c) shows a top image of a real sample; the scale bar represents 1 mm.

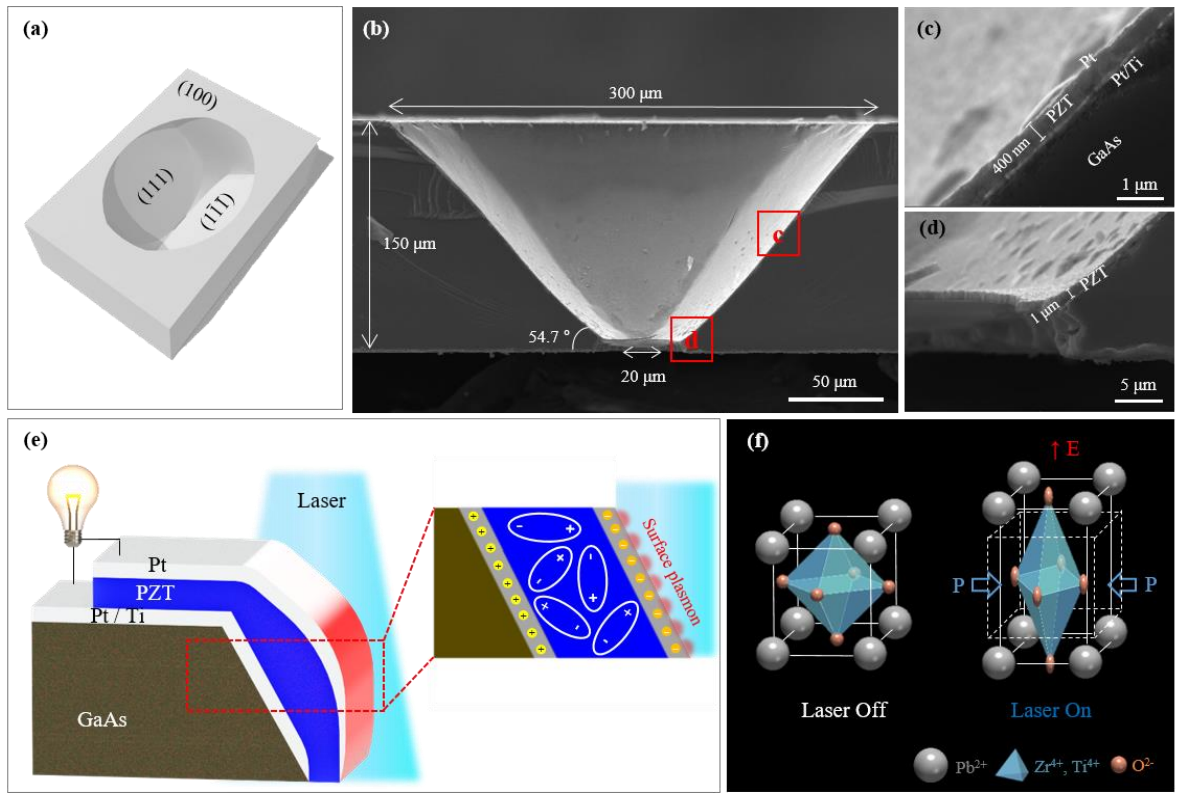

**Fig. 2 | Mechanism of the LPEG driven by light. a**, 3D illustration of the wet-chemical-etched GaAs formed to concentrate light. **b,** Cross-sectional FE-SEM image of etched GaAs. The upper and lower distances are 300 μm and 20 μm, respectively, and the thickness is 150 μm. **c**, Middle part of the slope in the microvoid (labelled 'c' in Fig. 2 (b)); the thickness of Pb(Zr,Ti)O$_3$ is 400 nm. Both sides of Pb(Zr,Ti)O$_3$ are covered with electrode. **d**, Lower part in the microvoid (labelled 'd' in Fig. 2 (b)); the thickness of Pb(Zr,Ti)O$_3$ is 1 μm thicker than in the upper part. **e**, Design and structure of the device. The GaAs microvoid structure has a Pb(Zr,Ti)O$_3$ thin film on the lower electrode (Pt/Ti) and the upper electrode is platinum. The electrodes are directly connected to an external load (left side). The light pressure increases due to the surface plasmon phenomenon in the area of laser illumination. Pb(Zr,Ti)O$_3$ induces polarization (right side). **f**, Atomic structure of the perovskite before and after laser irradiation.

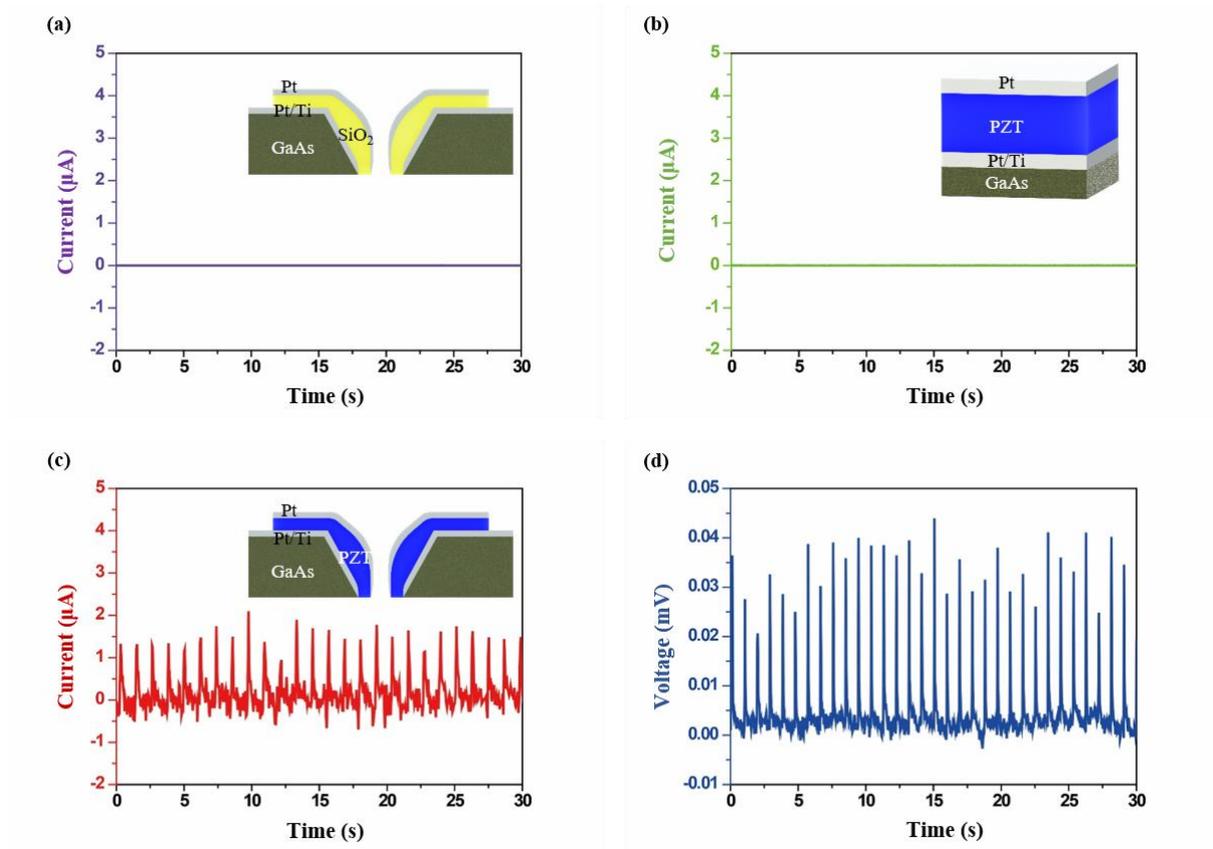

**Fig. 3 | Output of LPEGs with different materials and structures.** Each inset shows the structure. **a**, Device replacing the piezoelectric layer with SiO$_2$: no current under laser irradiation. **b**, Device based on the Pb(Zr,Ti)O$_3$ layer but with a flat structure: similarly no current response. **c** and **d**, Device based on the Pb(Zr,Ti)O$_3$ layer with a microvoid structure: large jumps in current and voltage occurred when the laser was turned on.

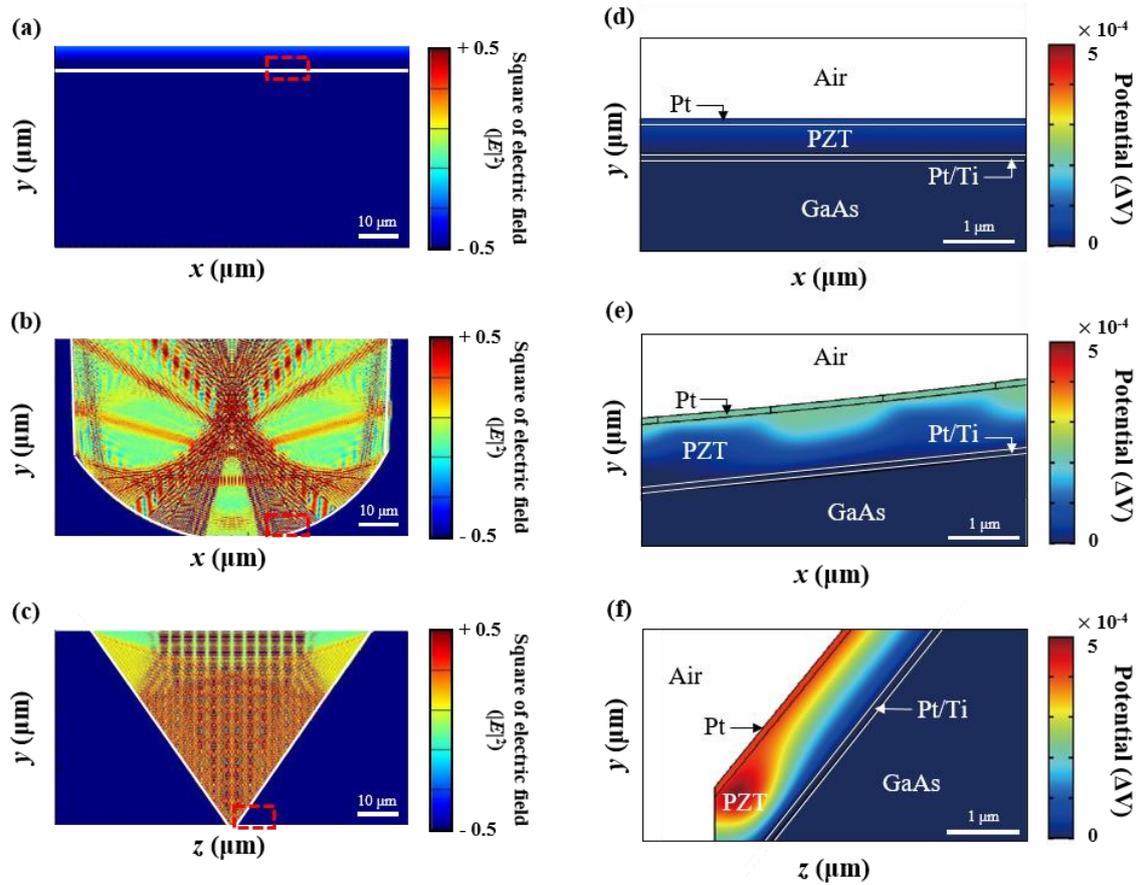

**Fig. 4 | FDTD and COMSOL simulations of electric field and potential distribution in flat and microvoid structures.** FDTD simulation results of the cross-sectional electric field intensity distribution when irradiated with 401 nm light for **a**, the $x$–$y$ plane in a flat structure; **b**, the $x$–$y$ plane in a microvoid structure; and **c**, the $z$–$y$ plane in the microvoid structure (logarithmic scale). The microvoid structure was modelled using 3D CAD, and the material of the simulated structure was platinum. COMSOL simulation results of cross-sectional potential distribution for **d**, the $x$–$y$ plane in the flat structure; **e**, the $x$–$y$ plane in the microvoid structure; and **f**, the $z$–$y$ plane in the microvoid structure for the red rectangle in (a)–(c), respectively. The microvoid structure was made based on the sample using 2D CAD.

**Acknowledgement**

We acknowledge financial support from the Basic Science Research Program through the National Research Foundation of Korea (NRF) funded by the Ministry of Education (No. 2017R1A2B4009832).

**Author contributions**

H. Y. Lee and S. N. Yi conceived the project; S. N. Yi and G.-T Hwang supervised it. H. Y. Lee fabricated the samples and performed FE–SEM, electric characteristics measurements, and FDTD simulation. H. Y. Lee and S. N. Yi performed the theoretical calculations and wrote most of the manuscript. K.-W. Lim performed COMSOL simulations. M. S. Kwak and H. Y. Lee drew the schematic diagrams. All authors contributed to improving the manuscript.

**Reprints and permissions information** is available at http://www.nature.com/reprints.

**Correspondence and requests for materials** should be addressed to S.N.Y.

*e-mail: snyi@kmou.ac.kr

[1]Department of Electronic Materials Engineering, Korea Maritime and Ocean University, Busan 49112, Korea

[2]Korea Institute of Materials Science (KIMS), Changwon, Gyeongnam, 51808, Korea

METHODS

**Device fabrication.** Wet-chemical etching was used to fabricate the GaAs microvoid structures[22]. Extended Data Fig. 7 presents the device manufacturing process. The photolithography pattern was formed using a 50 μm circular photomask. Each fabricated structure was covered with a Pt/Ti (100 nm / 10 nm) base electrode deposited by e-beam evaporation. Commercially available 0.4 M PZT solution (Quintess, Incheon, Korea) was used to fabricate a thin PZT film (Extended Data Fig. 8) on the Pt/Ti/etched GaAs (100) wafer. The PZT solution was spin coated at 3000 rpm for 45 s, followed by pyrolysis on a hot plate at 250 °C with a soak time of 5 min. The PZT film thickness was ~120 nm after a single deposition step; deposition was repeated four times resulting in a film thickness of over 400 nm showing reasonable resistance (minimum ~kΩ). The thin pyrolyzed PZT films underwent rapid thermal annealing at 650 °C for 5 min. Extended Data Fig. 9 presents the piezoelectric constant of the $Pb(Zr,Ti)O_3$ thin film. The platinum layer of the top electrode was then patterned with a shadow mask at a thickness of 100 nm using the same method as for the base electrode.

**Computational details.** The FDTD simulations considered the following conditions (Extended Data Fig. 3): the FDTD region, the electric-field monitor, the light source, the movie monitor, and the object. 1) The top and bottom layers of the FDTD region absorb light fully without reflection, meaning that they are perfectly matched layers, and an asymmetric boundary condition was applied because the structure is symmetrical about the $y$-axis. The simulation time of 10,000 fs was sufficient to simulate the exact plasmon phenomenon. 2) The electric-field monitor was set up to contain enough of the object, and mesh sizes of 1 nm × 1 nm (Fig. 3 (b), inset) and 7 nm × 7 nm (Fig. 3(c), inset) were applied. 3) A total-field scattered-field source of wavelength 401 nm propagated along the $y$-axis at the top of the object. 4) The movie monitor checked the path of light moving in the structure in real time and was set to include enough of the structure. 5) The GaAs microvoid structure was set to platinum, the material that meets the light in the actual sample. The COMSOL simulation considered the object and the pressure: 1) Objects were fabricated using 2D CAD, and each layer was set up identically to the real material; and 2) the pressure was converted based on the electric-field values in the FDTD simulation and applied differently for each part.

**Measurements.** Currents were measured using a 6485 picoammeter (Tektronix, Cleveland, United States) by blue laser (LS Korea, Incheon, Korea), and charge data were measured using a 5015A charge amplifier (Kistler, Winterthur, Swiss) by the same source. Voltage data were measured using a sourcemeter system (Tektronix, Cleveland, United States). The laser spot size was approximately $7.85 \times 10^{-7}$ m$^2$, and measurements were taken at room temperature. The thickness of the films was monitored by a Mira-3 FE–SEM (TESCAN, Brno, Czech) at 20 kV in cross-sectional view.

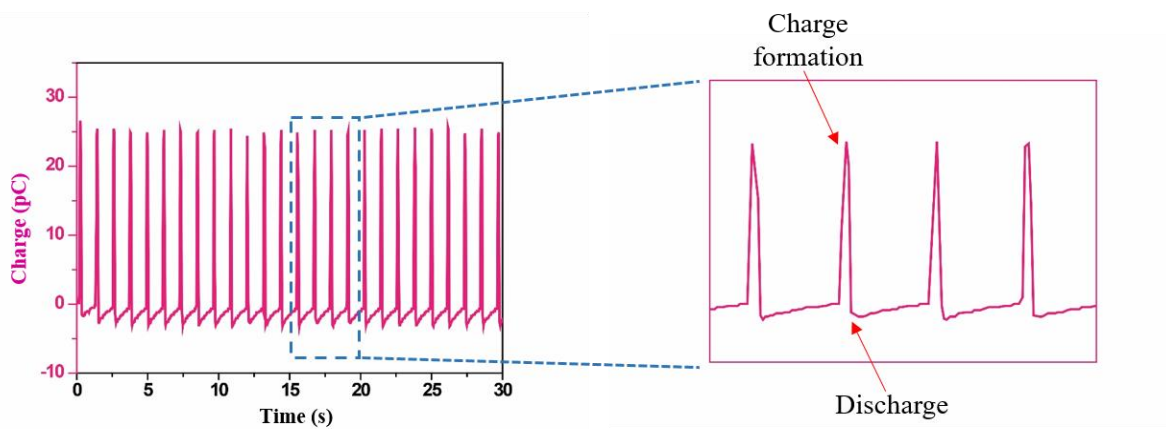

**Extended Data Fig. 1 | Charge measurement.** A large jump of 25 pC occurred in the light-pressure electric generator when a 10 mW (3.33 × $10^{-11}$ N) laser was turned on.

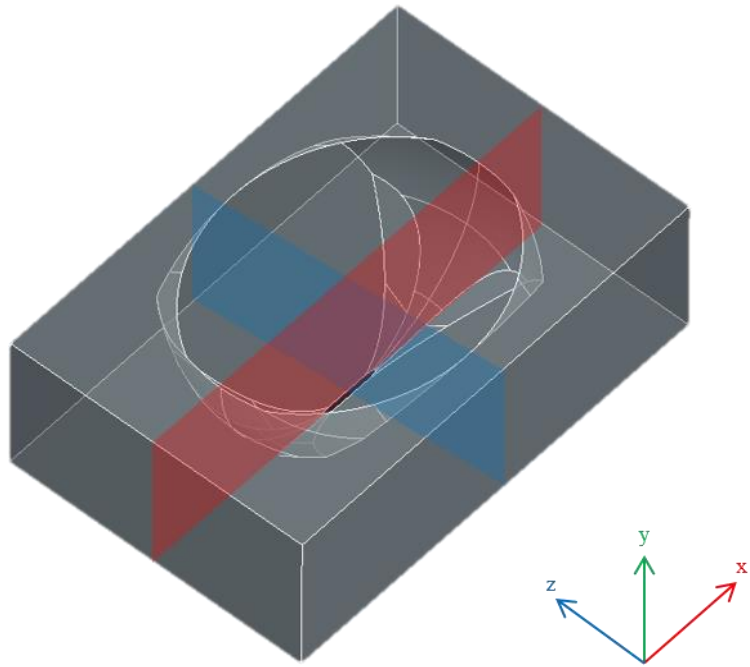

**Extended Data Fig. 2 | Three-dimensional GaAs microvoid structure drawn using CAD.**
The red and blue rectangles represent the x–y and y–z planes, respectively.

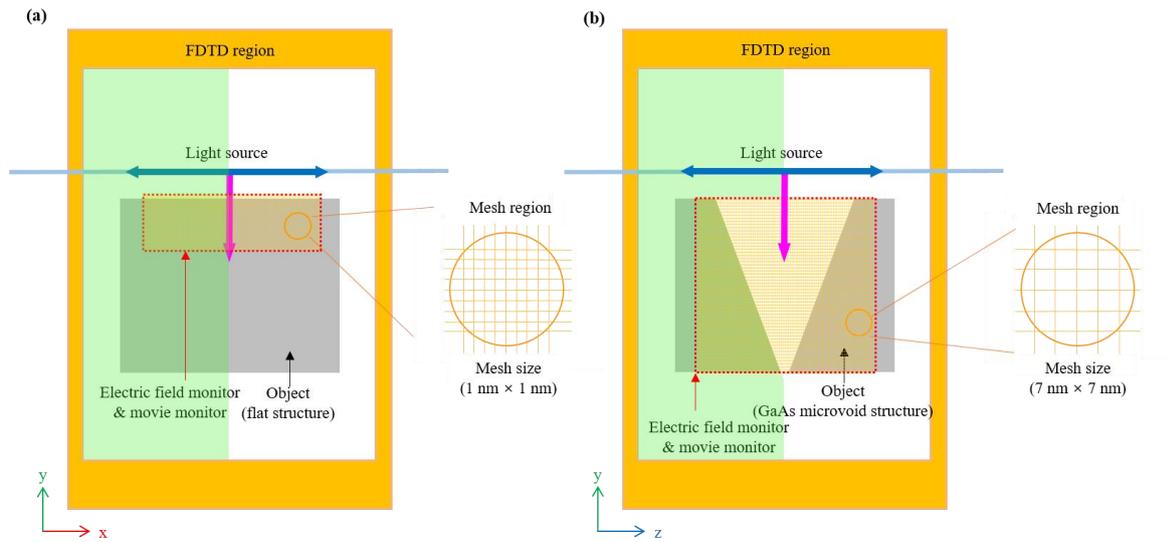

**Extended Data Fig. 3 | FDTD simulation conditions, including the FDTD region, electric-field monitor, movie monitor, light source, object, and mesh region. a,** flat structure; **b,** GaAs microvoid structure.

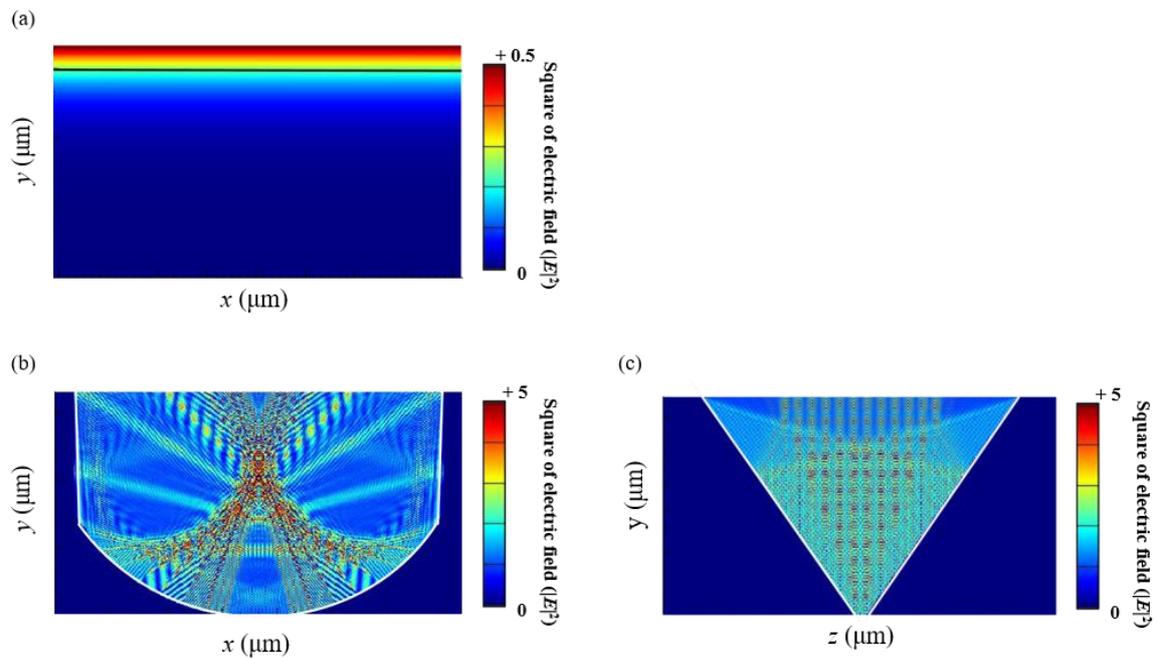

**Extended Data Fig. 4 | Simulated results for a square of electric field by radiation pressure in flat and microvoid structures.** FDTD simulation results for cross-sectional electric-field intensity distributions on a square of electric field scaled at 401 nm. The material is set as platinum. The microvoid structure was constructed using 3D CAD. **a**, *x–y* plane in the flat structure; **b**, *x–y* plane in the microvoid structure; **c**, *y–z* plane in the microvoid structure.

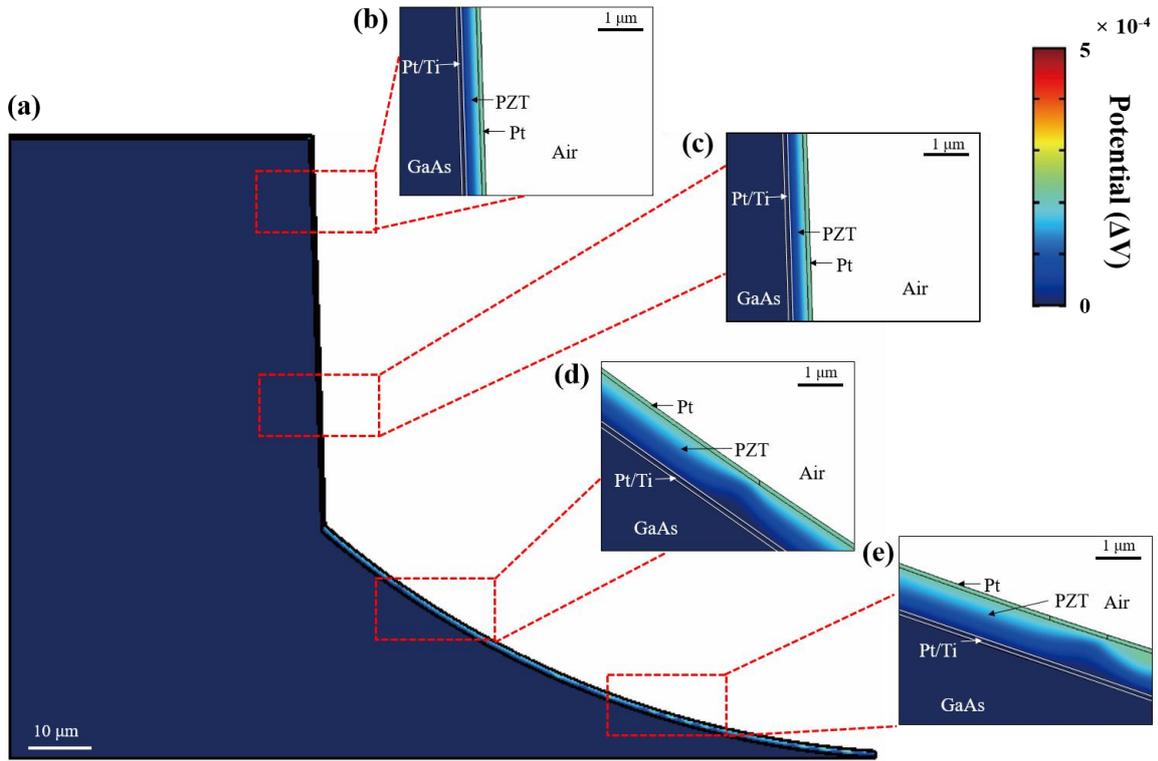

**Extended Data Fig. 5 | COMSOL simulation data of x–y plane. a,** overall diagram; **b,** first; **c,** second; **d,** third; **e,** fourth positions from top.

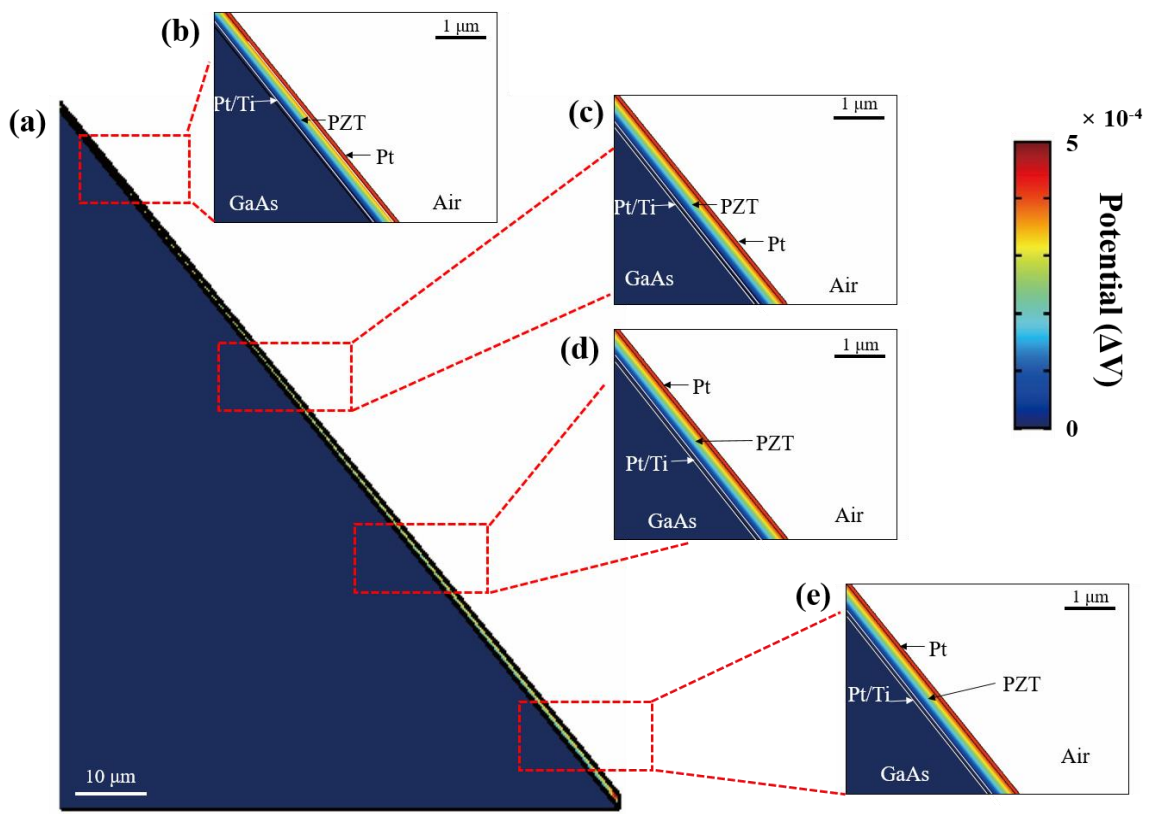

**Extended Data Fig. 6 | COMSOL simulation data of y–z plane. a,** overall diagram; **b–e,** first, second, third, and fourth positions from the top.

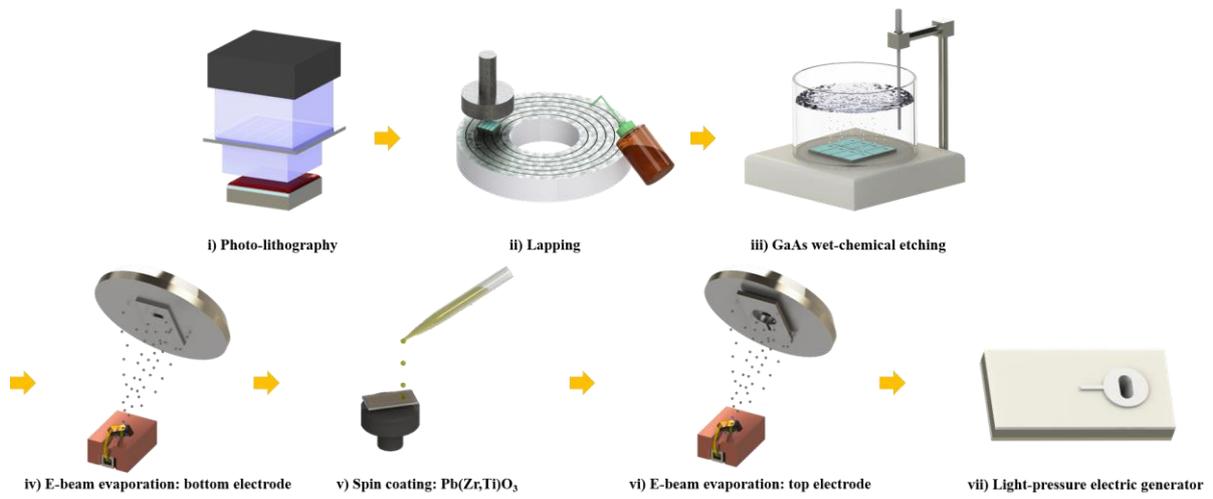

**Extended Data Fig. 7 | Device manufacturing process.**

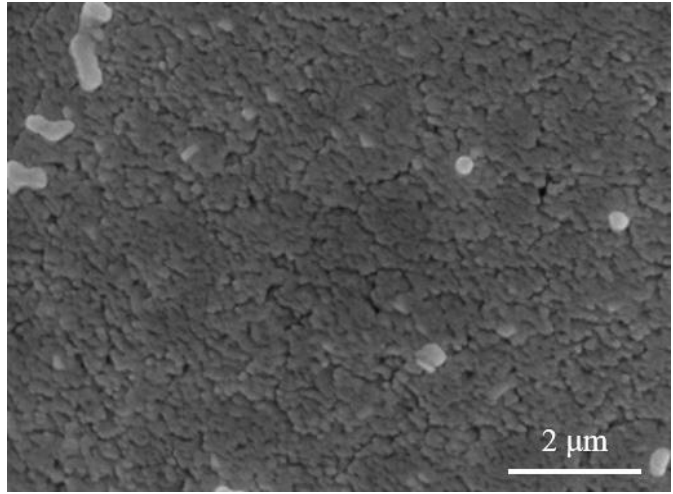

**Extended Data Fig. 8 | FE–SEM image of Pb(Zr,Ti)O$_3$ grains.**

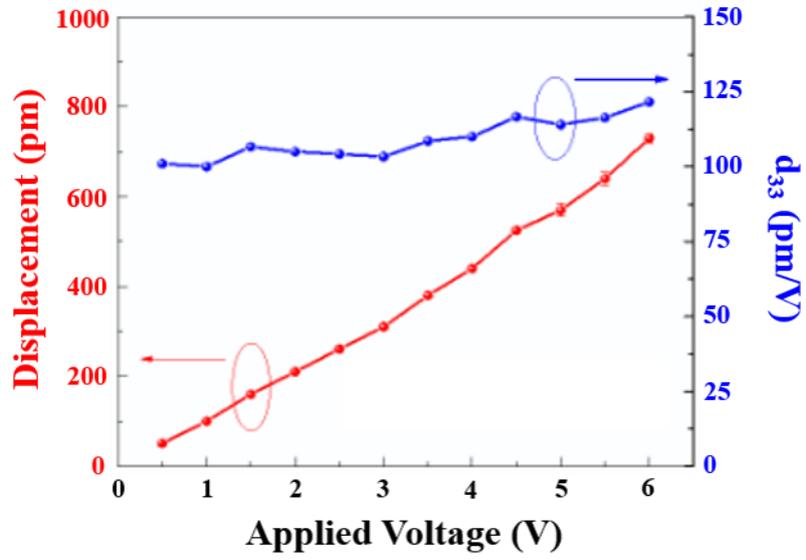

**Extended Data Fig. 9 | Piezoelectric constant d33 of Pb(Zr,Ti)O3 (52/48) for 1-μm-thick film.** The mean value of $d_{33}$ is 115 pm/V. (Data supplied by Quintess company.)

**Calculation of radiation pressure and electric field for a 10 mW laser**

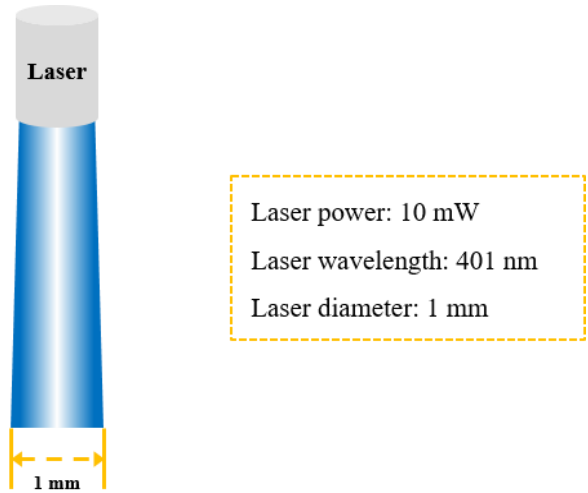

1. Calculation of radiation pressure from the laser specifications (*F*, force; *P*, power; *A*, area; *c*, light velocity; *r*, radius of laser).

$$F = \frac{P}{c} = \frac{1.0 \times 10^{-2} W}{299792458 m/s} = 3.33 \times 10^{-11} N$$

$$A = \pi r^2 = \pi \left(\frac{1.0 \times 10^{-3}}{2}\right)^2 = 7.85 \times 10^{-7} m^2 \; (r = 0.5 \text{ mm})$$

$$P_{ressure} = \frac{F}{A} = \frac{3.33 \times 10^{-11} N}{7.85 \times 10^{-7} m^2} = 4.24 \times 10^{-5} Pa \; (N/m^2).$$

2. Expression of radiation pressure, *P*, in terms of electric field (*I*, intensity; *E*, electric field; $\mu_0$, permeability of free space).

$$P = \frac{2I}{c} = \frac{2}{c^2 \mu_0} E^2 = 1.769 \times 10^{-11} E^2.$$

3. Calculation of electric field (combining equation systems 1 and 2).

$$1.769 \times 10^{-11} E^2 = 4.24 \times 10^{-5} Pa$$

$$\therefore E^2 = 2.39 \times 10^6 \; V/m \; (\text{or } E = 1546 \text{ V/m}).$$

**An example of inferring radiation pressure based on FDTD simulation results**

1. When the laser (power 10 mW, pressure 4.24 × 10$^{-5}$ Pa) was shone on the flat surface, and the power intensity, $I$, at a point was reduced to 0.25 of its original value according to the FDTD results, the pressure, $P$, at that point was

$$P = \frac{2I}{c} = \frac{2}{c^2\mu_0}xE^2 = 1.769 \times 10^{-11} \times 0.25 \times 2.39 \times 10^6 \, V/m = 1.05 \times 10^{-5} \, \text{Pa},$$

where $c$ is the velocity of light, $E$ is the electric field, $\mu_0$ is the permeability of free space, and $x$ is the increase factor of the electric field.

2. Irradiating the same laser (power, 10 mW; pressure, $P$, 4.24 × 10$^{-5}$ Pa) on a flat surface with its power intensity at a point increased to four times the original according to the FDTD results gives the pressure at the point as

$$P = \frac{2I}{c} = \frac{2}{c^2\mu_0}xE^2 = 1.769 \times 10^{-11} \times 4 \times 2.39 \times 10^6 \, V/m = 1.69 \times 10^{-4} \, \text{Pa},$$

where $I$ is intensity, $c$ is the velocity of light velocity, $E$ is the electric field, $\mu_0$ is the permeability of free space, and $x$ is the increase factor of the electric field.